\documentclass[conference]{IEEEtran}
\IEEEoverridecommandlockouts

\usepackage{etex}
\usepackage{amsmath,amssymb}
\usepackage{float}
\usepackage{lipsum}
\usepackage{array}
\usepackage{cite}
\usepackage{textcomp}
\usepackage{hhline}
\usepackage{siunitx}
\usepackage{balance}
\usepackage{makecell}
\usepackage{float}
\usepackage[pdftex]{graphicx}
\usepackage{siunitx}
\usepackage[dvipsnames]{xcolor}
\usepackage{xcolor}
\usepackage{colortbl}
\usepackage{tikz}
\usetikzlibrary{decorations}
\usetikzlibrary{calc} 
\tikzset{>=latex}
\usetikzlibrary{arrows,shapes}
\usetikzlibrary{arrows}
\usetikzlibrary{plotmarks}
\usepackage{ctable}
\usepackage{pgfplots}
\usepackage[english]{babel}
\usepackage{color}
\usepackage[utf8]{inputenc}
\usepackage[T1]{fontenc}
\usetikzlibrary{decorations.pathreplacing,shapes.misc}
\usetikzlibrary{patterns}
\usepackage{tcolorbox}
\usepackage{balance}
\usepackage{subfigure}
\usepackage{hyphenat}
\usepackage{amsmath,bm}

\usepackage{amsthm}
\theoremstyle{definition}

\addto\captionsenglish{}

\newcommand{\mL}{\mathcal{L}}

\newcommand{\nc}{n}

\newcommand{\kc}{k}

\newcommand{\lr}{\boldsymbol{r}}

\newcommand{\lrt}{\tilde{\boldsymbol{r}}}
\newcommand{\lc}{\boldsymbol{c}}
\newcommand{\hlc}{\hat{\boldsymbol{c}}}
\newcommand{\llr}{\boldsymbol{l}}
\newcommand{\rr}{\boldsymbol{R}}

\newcommand{\cc}{\boldsymbol{C}}

\newcommand{\dgmd}{\mathsf{d}_\mathsf{GD}}
\newcommand{\dmin}{d_\mathsf{min}}

\newcommand{\lalone}{\boldsymbol{L}}

\newcommand{\w}{w}
\newcommand{\BB}{\mathsf{B}}

\newcommand{\argmin}[1]{\underset{#1}{\mathrm{arg \, min}}\,}

\newcommand{\Eb}{\mathsf{E_b}}
\newcommand{\No}{\mathsf{N_0}}

\def\forcemath#1{\ifmmode #1 \else $#1$\fi}

\newcommand{\ham}{\mathsf{d}_\mathsf{H}}

\ifCLASSINFOpdf
 \else
  \fi

\begin{document}


\title{Binary Message Passing Decoding of Product Codes Based on Generalized Minimum Distance Decoding}

\author{Alireza Sheikh$^\mathsection$, Alexandre Graell i
Amat$^\mathsection$, and Gianluigi Liva$^\dagger$ \\ \IEEEauthorblockA{$^\mathsection$ Department of Electrical Engineering, Chalmers University of Technology, Sweden \\ $^\dagger$Institute of Communications and Navigation of the German Aerospace Center (DLR), Germany \\~\\ (Invited Paper)
\thanks{This work was financially supported by the Knut and Alice
Wallenberg Foundation, the Swedish Research Council under grant 2016-04253,
and the Ericsson Research Foundation.}
}}


\maketitle

\IEEEpeerreviewmaketitle

\begin{abstract}
We propose a binary message passing decoding algorithm for product codes based on generalized minimum distance decoding (GMDD) of the component codes, where the last stage of the GMDD makes a decision based on the Hamming distance metric. The proposed algorithm closes half of the gap between conventional iterative bounded distance decoding (iBDD) and turbo product decoding based on the Chase--Pyndiah algorithm, at the expense of some increase in complexity. Furthermore, the proposed algorithm entails only a limited increase in data flow compared to iBDD. 
\end{abstract}

\section{Introduction}

Applications requiring very high throughputs, such as fiber-optic communications and high-speed wireless communications, have recently triggered a significant amount of research on low-complexity decoders. While codes-on-graphs such as low-density parity-check (LDPC) codes and turbo codes have been shown to provide close-to-capacity performance under belief propagation (BP) decoding, scaling their BP decoders to yield throughtputs of the order of several Gbps or even  $1$ Tbps, as required for example for the the future optical metro-networks, is a very challenging task. One of the main bottlenecks is the data flow required by the exchange of soft information in the iterative BP decoding. This has spurred a great deal of research in novel low-complexity decoding algorithms. 

Several works have attempted to reduce the decoding complexity of BP decoding of LDPC codes, see, e.g., \cite{Dar10, Moh10, Ang14,Cus16}. For high-throughput applications, an alternative to LDPC codes with (BP) soft decision decoding (SDD)  is to consider hard decision decoding (HDD). Product codes (PCs)\cite{Eli54}, half-product codes \cite{Justesen2011}, staircase codes \cite{staircase_frank}, braided codes \cite{Jia17}, and other product-like code structures \cite{Haeger2017tit} with HDD based on bounded distance decoding (BDD) of the component codes (which we refer here to as iterative BDD (iBDD)) yield excellent performance with a  significantly reduced data flow, hence achieving very high throughputs. However, this comes at the expense of a performance loss (typically larger than 1 dB) compared to SDD.

To close the performance gap between iBDD of product-like codes and SDD of LDPC codes or product-like codes, yet with throughputs and energy consumption close to that of iBDD, another line of research recently explored is to improve the performance of the conventional iBDD. In \cite{Hag18}, an algorithm that exploits conflicts between
component codes  in order to assess their reliabilities even when no
channel reliability information is available, was proposed. The algorithm, dubbed anchor decoding (AD), improves the performance of iBDD at the expense of some increase in decoding complexity. In \cite{Yibitflip}, a decoding algorithm based on marking the least reliable bits was proposed for staircase codes. In \cite{She18}, we proposed a decoding algorithm based on BDD of the component codes, named iBDD with scaled reliability (iBDD-SR). The
algorithm in \cite{She18} improves the performance of iBDD by exploiting channel reliabilities as proposed in \cite{lechner2012analysis} for LDPC codes, while still only exchanging binary (i.e., hard-decision)
messages between component decoders, similar to iBDD. iBDD-SR improves upon iBDD and AD, and achieves the same throughput of iBDD with a slight increase in energy consumption \cite{Fou19}. In \cite{She18b}, we proposed an algorithm based on generalized minimum distance  decoding (GMDD)  of the component codes. The proposed algorithm closes over $50\%$ of the performance gap between iBDD and turbo product decoding (TPD) based on the Chase--Pyndiah algorithm \cite{Pyn98}, with lower complexity than TPD. However, the algorithm, which we referred to as iterative GMDD with scaled reliability (iGMDD-SR), requires the exchange of soft information between the component decoders and hence entails a decoder data flow equivalent to that of TPD and significantly higher than that of iBDD.

In this paper, we propose a novel binary message passing (BMP) decoding algorithm for product codes based on  GMDD of the component codes, which we refer to as BMP-GMDD. The proposed algorithm follows the same principle as the iGMDD-SR algorithm proposed in \cite{She18b}, but a crucial difference is that the Hamming distance metric is used at the final stage of the GMDD of the component codes. In contrast to iGMDD-SR, the resulting algorithm does not require the exchange of soft information, but the exchange of the hard decisions on the code bits (as conventional iBDD) and an ordered list of the $\dmin-1$ least reliable code bits for each component code, where $\dmin$ is the minimum Hamming distance of the component code. This list can be represented by a small number of bits. The proposed algorithm yields performance very close to that of iGMDD-SR, closing $50\%$ of the performance gap between iBDD and TPD, while entailing only a small increase in data flow compared to iBDD (between $8.5\%$ and $34.3\%$, depending on the code parameters).

\emph{Notation:} We use boldface letters to denote vectors
and matrices, e.g., $\boldsymbol{x}$  and $\boldsymbol{X}=[x_{i,j}]$. The $i$-th
row and $j$-th column of matrix $\boldsymbol{X}$ is
denoted by $\boldsymbol{X}_{i,:}$ and $\boldsymbol{X}_{:,j}$, respectively. 
$|a|$ denotes the absolute value of $a$,
$\left\lfloor a \right\rfloor$ the largest integer smaller than or
equal to $a$, and $\left\lceil a \right\rceil$ the smallest integer larger than or
equal to $a$. 
A Gaussian distribution with mean $\mu$ and variance $\sigma^2$ is denoted by $\mathcal{N}(\mu ,\sigma^2)$. 

\section{Preliminaries}
\label{sys_mod} 

\begin{figure}[!t]
	\centering
	\includegraphics[width=3.55cm]{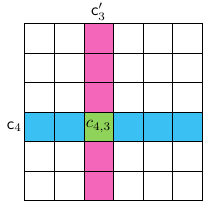}
	$\qquad$
	\includegraphics[scale=1.12]{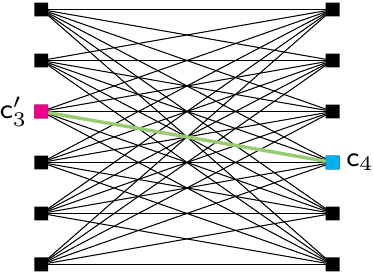}
	\caption{Code array (left) and simplified Tanner graph (right) of a PC with identical 
	component code of length $\nc=6$ for row and column codes. In the simplified Tanner graph, the CNs are represented by squares (the CNs on the left represent the column codes and the CNs on the right represent the row codes) and 
	degree-$2$ VNs are represented as simple edges. The third column code and the fourth row code are highlighted.}
	\label{fig:PCSimpGraph}
	 \vspace{-3mm}
\end{figure}

Let $\mathcal{C}$ be an $(\nc, \kc, \dmin)$ binary linear  code, where
$\nc$, $\kc$, and $\dmin$ are the code length, dimension, and minimum
distance, respectively. We consider two-dimensional PCs with identical binary Bose--Chaudhuri--Hocquenghem (BCH) component code $\mathcal{C}$ for the row and column codes. Such a PC, of parameters
$(\nc^2,\kc^2,\dmin^2)$ and rate $R = \kc^2/\nc^2$, is defined as the
set of all $\nc\times\nc$ arrays such that each row and each column of the
array is a codeword of $\mathcal{C}$. Thus, a codeword of the
PC can be represented by an $\nc\times \nc$ binary matrix $\cc=[c_{i,j}]$. Alternatively, a PC can be defined via a
Tanner graph with $2\nc$ constraint nodes (CNs), where $\nc$ CNs
correspond to the row codes and $\nc$ CNs correspond to the column
codes. The graph has $\nc^2$ variable nodes (VNs) corresponding to the
$\nc^2$ code bits.  The code array and (simplified) Tanner graph of a
two-dimensional PC with $\nc=6$ is shown in Fig.~\ref{fig:PCSimpGraph}.

We assume transmission over the binary-input additive white Gaussian
noise (AWGN) channel. The
channel observation corresponding to code bit $c_{i,j}$ is given by
\begin{align*}
y_{i,j}=x_{i,j}+z_{i,j},
\end{align*}  
where $x_{i,j}=(-1)^{c_{i,j}}$, $z_{i,j}\sim \mathcal{N}(0,(2 R
\Eb/\No)^{-1})$, with $\Eb/\No$ being the signal to noise ratio. We denote by $\lalone=[L_{i,j}]$ the matrix of channel
log-likelihood ratios (LLRs) and by $\rr=[r_{i,j}]$ the matrix of hard
decisions at the channel output, where $r_{i,j}$ is obtained by
mapping the sign of $L_{i,j}$ according to $ 1 \mapsto 0$ and $- 1
\mapsto 1$. We denote this mapping by $\BB(\cdot)$, i.e.,
$r_{i,j}=\BB(L_{i,j})$. With some abuse of notation, we also write
$\rr=\BB(\lalone)$.

\subsection{Generalized Minimum Distance Decoding}
\label{sec:GMD}

Consider the decoding of a BCH component code of length $\nc$ and the vector of channel LLRs $\llr=(L_1,\ldots,L_{\nc})$ corresponding to the received vector $\lr=(r_1,\ldots,r_\nc)$. GMDD is based on multiple algebraic error-erasure decoding attempts \cite{ForneyGMD}.  In
particular, the decoder ranks the coded bits in terms of their
reliabilities $|L_1|,\ldots,|L_{\nc}|$. Then, the $m$ least reliable
bits in $\lr$ are erased, where $m \in \mathcal{M}_\mathsf{odd}\triangleq\{\dmin-1,\dmin-3,...,2\}$ if
$\dmin$ is odd and $m \in \mathcal{M}_\mathsf{even}\triangleq \in \{\dmin-1,\dmin-3,...,3\}$ if $\dmin$ is
even. For later use, we denote by $\mL$ the ordered list of $\dmin-1$ least reliable code bits. It can be readily checked that $|\mathcal{M}_\mathsf{odd}|=|\mathcal{M}_\mathsf{even}|=t$, where $t = \left\lfloor \frac{\dmin-1}{2} \right\rfloor$ is the error correcting capability of the code. Together with the received vector $\lr$, this gives a list of $t+1$ trial vectors
$\lrt_{i}$, $i = 1, \dots,
t+1$, out of which $t$ vectors contain both erasures and
(possibly) errors. Finally, algebraic error-erasure decoding
\cite[Sec.~6.6]{LinCos04} is applied to each trial vector $\lrt_i$, resulting in a set of candidate codewords, of size at most $t+1$, denoted by $\mathcal{S}$ .
If decoding fails for all $t+1$ vectors in the list, a failure is declared. Otherwise, the
decoder picks among all candidate codewords in $\mathcal{S}$ the
one that minimizes the generalized distance $\dgmd(\lr,\lc)$, \cite{ForneyGMD}
\begin{align}\label{GMDmetric}
\hlc&=\argmin{\lc\in\mathcal{S}}\dgmd(\lr,\lc)\nonumber\\
&=\argmin{\lc\in\mathcal{S}}\sum\limits_{i:{r_{i}} = {c_{i}}} {\left( {1 - {\alpha_{i}}} \right)}  + \sum\limits_{i:{r_{i}} \ne {c_{i}}} {\left( {1 + {\alpha_{i}}} \right)}, 
\end{align}
where $\alpha_{i}\buildrel \Delta \over = |L_{i}|/\mathop {\max
}\limits_{1 \le j \le \nc} |L_{j}|$. Note that if all input LLRs
$L_{1},\ldots,L_{n}$ have the same magnitude, we have $\alpha_i = 1$ for all $i =
1,\dots,\nc$ and \eqref{GMDmetric} reverts to $2 \ham(\lr, \hlc)$, where $\ham(\lr, \hlc)$ is the Hamming distance between $\lr$ and $\hlc$.

By introducing erasures and performing multiple error-erasure component
decoding attempts, GMDD can decode beyond half the minimum distance of the code.



\section{Binary Message Passing Decoding based on Generalized Minimum Distance Decoding}\label{iHMDDSR}

In this section, we propose a BMP decoding algorithm for
PCs based on GMDD of the component codes. We refer to it as BMP-GMDD. The algorithm follows the same principle as the iGMDD-SR algorithm that we proposed in \cite{She18b}. However, compared to iGMDD-SR, the proposed BMP-GMDD does not require the exchange of the reliabilities on the code bits between the row and column decoders. To achieve that, rather than considering the generalized distance in \eqref{GMDmetric} to perform the decision at the last stage of GMDD of the row and column decoders as in \cite{She18b}, we perform the decision based on the Hamming distance, i.e., among all candidate codewords in $\mathcal{S}$ (see Section~\ref{sec:GMD}), the decoder selects the one that minimizes $\ham(\lr,\lc)$, i.e., the decision in \eqref{GMDmetric} is substituted by
\begin{align}\label{HDmetric}
\hlc=\argmin{\lc\in\mathcal{S}}\ham(\lr,\lc).
\end{align} 

Making the decision based on the Hamming distance instead of the generalized distance entails a small performance loss, as the decision does not take into consideration the normalized reliabilities $\alpha_i$. However, this allows to significantly reduce the decoder data flow, as explained later.

The proposed BMP-GMDD algorithm works as follows. Without loss of generality, assume that the
decoding starts with the row codes and let us consider the decoding of
the $i$-th row code at iteration $\ell$.  Let $\bm{\Psi}^{\mathsf{c},(\ell-1)}=[\psi_{i,j}^{\mathsf{c},(\ell-1)}]$ be the matrix of hard decisions on code bits $c_{i,j}$ after the decoding of the $\nc$ column codes at iteration $\ell-1$. Also, let $\mL_i^{\mathsf{r},(\ell-1)}$ be  the ordered list of $\dmin-1$ least reliable bits of codeword $\cc_{i,:}$ from the decoding of the column codes at iteration $\ell-1$. Note that in the first iteration the list $\mL_i^{\mathsf{r},(\ell-1)}$ is built according to the ordering of the channel reliabilities $\lalone_{i,:}=(L_{i,1},\ldots,L_{i,\nc})$. Row decoding of the $i$-th row code is then performed based on $\bm{\Psi}_{i,:}^{\mathsf{c},(\ell-1)}$ and $\mL_i^{\mathsf{r},(\ell-1)}$. First, GMDD of the $i$-th row code based on the Hamming distance is performed based on $\bm{\Psi}^{\mathsf{c},(\ell-1)}_{i,:}$ and  $\mL_i^{\mathsf{r}}$,  as explained in Section~\ref{sec:GMD} (see \eqref{HDmetric}). Note that GMDD does not provide reliability information about the decoded
bits, i.e., it is a \emph{soft-input} \emph{hard-output} decoding algorithm. In order to provide the column decoders with the list of $m$ least reliable bits for each codeword $\cc_{:,j}$ after the decoding of the row codes at iteration $\ell$, we do the following. The output bits of GMDD are mapped according to
$0 \mapsto +1$ and $1 \mapsto -1$ if GMDD is successful and mapped to $0$
if GMDD fails. Let $\bar{\mu}_{i,j}^{\mathsf r, (\ell)} \in
\{\pm1, 0 \}$ be the result of this mapping for the decoded bit corresponding to code bit $c_{i,j}$. The reliability information is then formed
according to
\begin{equation}\label{eq:GMDchrel_VN_scale}
\mu_{i,j}^{\mathsf r, (\ell)}=\w_i^{\mathsf r, (\ell)} \cdot \bar{\mu}_{i,j}^{\mathsf r,
(\ell)} + L_{i,j}, 
\end{equation}
where $\w_i^{\mathsf r, (\ell)} > 0$ is a scaling factor than needs to be optimized. 
Then, the hard decision on $c_{i,j}$ made by the $i$-th row decoder is formed as
\begin{equation}\label{eq:BDDchrel_VN_scale}	
\psi_{i,j}^{\mathsf{r},(\ell)}=
\BB(\mu_{i,j}^{\mathsf r, (\ell)}).
\end{equation}

The hard decision $\psi_{i,j}^{\mathsf{r},(\ell)}$ is the binary message on code bit $c_{i,j}$
passed from the $i$-th row code to the $j$-th column code, i.e., from the $i$-th row CN to the $j$-th column CN (see Fig.~\ref{fig:PCSimpGraph}). In
particular, after applying this procedure to all row codes, the matrix
$\boldsymbol{\Psi}^{\mathsf{r},(\ell)}=[\psi_{i,j}^{\mathsf{r},(\ell)}]$
is formed and used as the input for the $n$ column decoders.  Furthermore, after decoding of all row codes, for each column codeword $\cc_{:,j}$, the corresponding code bits are ranked according to the reliabilities $(\mu_{1,j}^{\mathsf{r},(\ell)},\ldots,\mu_{\nc,j}^{\mathsf{r},(\ell)})$. Then the $m$ least reliable bits are stored in the list $\mL_j^{\mathsf{c},(\ell)}$, which is passed to the $j$-th column decoder.

\begin{figure}[t!]
	\centering
	\includegraphics[width=\columnwidth]{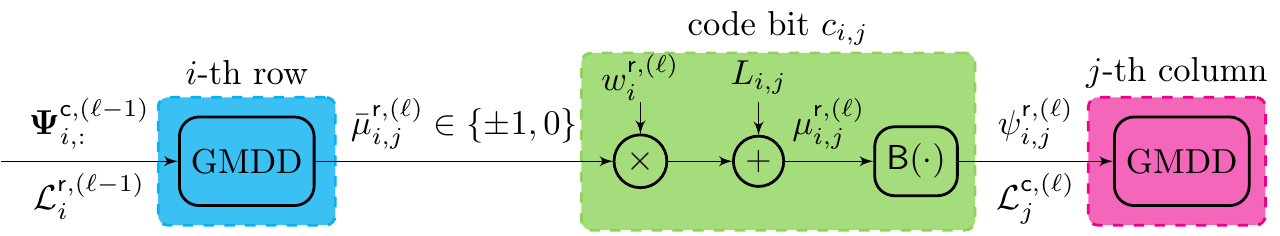}
	\caption{Block diagram showing the information flow from the $i$-th row decoder to
	the $j$-th column decoder in BMP-GMDD. The message at the input of the $i$-th row decoder is the vector of hard decisions on the code bits $\bm{\Psi}^{\mathsf{c},(\ell-1)}_{i,:}$ and the ordered list of the $\dmin-1$ least reliable bits $\mL_j^{\mathsf{r},(\ell-1)}$ from the decoding of the column codes at the previous iteration.}
	\label{messagepasing2}
\end{figure}

The decoding of the $\nc$ column codes at iteration $\ell$ is then performed based on the hard decisions $\boldsymbol{\Psi}^{\mathsf{r},(\ell)}$ and the lists of least reliable bits $\mL_1^{\mathsf{c},(\ell)},\ldots,\mL_\nc^{\mathsf{c},(\ell)}$ as explained for the $i$-th row decoder above. After decoding of the $\nc$ column codes at decoding iteration $\ell$, the matrix $\bm{\Psi}^{\mathsf{c},(\ell)}=[\psi_{i,j}^{\mathsf{c},(\ell)}]$ of hard decision bits and the lists $\mL_1^{\mathsf{r},(\ell)},\ldots,\mL_\nc^{\mathsf{r},(\ell)}$ are passed to the $\nc$ row decoders for the next decoding iteration. The iterative process continues until a maximum number of
iterations is reached.   The BMP-GMDD of PCs is schematized in Fig.~\ref{messagepasing2}.

Remark: With reference to Fig.~\ref{messagepasing2}, the iGMDD-SR algorithm proposed in \cite{She18b} passes the soft information $\mu_{i,j}^{\mathsf r, (\ell)}$ to the $j$-th column decoder, which entails a significantly higher decoder data flow compared to BMP-GMDD.

\section{Decoding Complexity Discussion}
\label{complexity}

A thorough complexity analysis of BMP-GMDD should include, besides pure algorithmic
aspects, implementation implications in terms of memory requirements, wiring, and transistor switching activity \cite{Fou19}, and is  beyond the scope of this paper. We
however provide a high-level discussion of the complexity and data flow of BMP-GMDD compared to that of conventional iBDD, AD \cite{Hag18}, iBDD-SR \cite{She18}, and iGMDD-SR \cite{She18b}. 

Conventional iBDD, iBDD-SR, and AD are based on BDD of the component codes and are characterized by a similar complexity and data flow. In particular, it was shown in \cite{Fou19} that for the same data throughput (up to 1 Tbps), iBDD-SR provides $0.2$--$0.25$ dB gain with respect to iBDD with only slightly higher energy consumption. 

Both GMDD-SR and the proposed BMP-GMDD are based on GMDD of the component codes.  In this case, $t$ error-erasure decoding attempts and one BDD attempt
are required. Each error-erasure decoding attempt has a cost  close to a run of BDD. Each decoding attempt may result in a candidate codeword that is used to
form a list of size up to $t+1$, as explained in Section~\ref{sec:GMD}. The minimization of the generalized distance in
\eqref{GMDmetric} for GMDD-SR and the Hamming distance in \eqref{HDmetric} for BMP-GMDD has a negligible cost with respect to the $t+1$ decoding attempts.  On the other hand, both BMP-GMDD and iGMDD-SR require finding the $\dmin-1$ least reliable bits and sorting them according to their reliabilities, which adds some further complexity.   
\newcommand{\tablehighlight}{}
\begin{table*}[t]
 \caption{Comparison of different product decoding algorithms. Coding
	gains and capacity gaps are measured at a BER of $10^{-6}$}
  \centering
  \renewcommand{\arraystretch}{1}
  \begin{tabular}{cccccc}
  		\toprule
		\tablehighlight{acronym} & \tablehighlight{decoding algorithm} &
		\makecell{\tablehighlight{channel}\\ \tablehighlight{reliabilities}} &
		\makecell{\tablehighlight{exchanged} \\
		\tablehighlight{messages}} & \makecell{gain over\\ iBDD (dB)}&
		\makecell{gap from\\ capacity (dB)}\\
		\midrule
		iBDD & iterative bounded distance decoding & no & hard & - & $1.03$ (HD) \\
		iBDD (ideal) & iterative bounded distance decoding without
        miscorrections & no & hard & $0.28$ & $0.75$ (HD) \\	
        iBDD-SR & iterative bounded distance decoding with scaled
        reliability \cite{She18} & yes & hard  & $0.27$ & $2.3$ (SD) \\		
		AD & anchor decoding \cite{Hag18} & no & hard & $0.18$ & $0.85$ (HD) \\
		BMP-GMDD & binary message passing decoding based on GMD decoding & yes & hard & $0.51$ & $1.79$ (SD) \\			
		iGMDD-SR & iterative generalized minimum distance decoding with
		scaled reliability \cite{She18b} & yes & soft & $0.58$ & $1.72$ (SD) \\
		TPD & turbo product decoding (Chase--Pyndiah) \cite{Pyn98} & yes & soft & $1.08$ & $1.22$ (SD) \\
		\bottomrule
  \end{tabular}
	\label{tab:Table1}
\end{table*}

Note that GMDD-SR requires the component decoders to be provided with soft information by the previous decoding iteration. Therefore, its data flow is significantly higher than that of iBDD, iBDD-SR, and AD, and is the same of soft decision TPD. For an $a$-bit representation of the soft information, the data flow is roughly $a$ times that of BDD, iBDD-SR, and AD. In contrast, BMP-GMDD requires only the exchange of the hard decisions and the ordered list of $\dmin-1$ least reliable bits for each row and column codeword.  For a component code of length $n$, the index of each code bit can be represented with $\left\lceil \text{log}_{2}(n) \right\rceil$ bits. Furthermore, for each of the $\dmin-1$ least reliable bits we need to provide their ordering in terms of reliabilities. Thus, each ordered lists $\mL_i^{\mathsf{r},(\ell)}$, $i=1,\ldots,\nc$, and $\mL_j^{\mathsf{c},(\ell)}$, $=j,\ldots,\nc$, can be represented with 
\begin{align*}
\left(\left\lceil\text{log}_{2}(n)\right\rceil + \left\lceil \text{log}_{2}(\dmin-1) \right\rceil\right) (\dmin-1)
\end{align*}
bits each. This is the additional data flow (per row and column code decoding) compared to conventional iBDD. For instance, for a component code of code length $\nc=256$ bits, the data flow of BMP-GDD is $15.625\%$ and $34.375\%$ higher than that of iBDD for $\dmin=5$ ($t=2$) and $\dmin=9$ ($t=4$), respectively. For a component code of length $\nc=512$ bits, the increase in data flow is reduced to $8.593\%$ and $18.75\%$, respectively. Thus, the increase in data flow of BMP-GMDD compared to iBDD is very limited and is much lower than the data flow of iGMDD-SR and conventional TPD.

\section{Numerical Results}

\renewcommand{\w}{\boldsymbol{w}}
\newcommand{\maxIter}{\ell_{\text{max}}}

\begin{figure}[t] \centering 
	\includegraphics[width=\columnwidth]{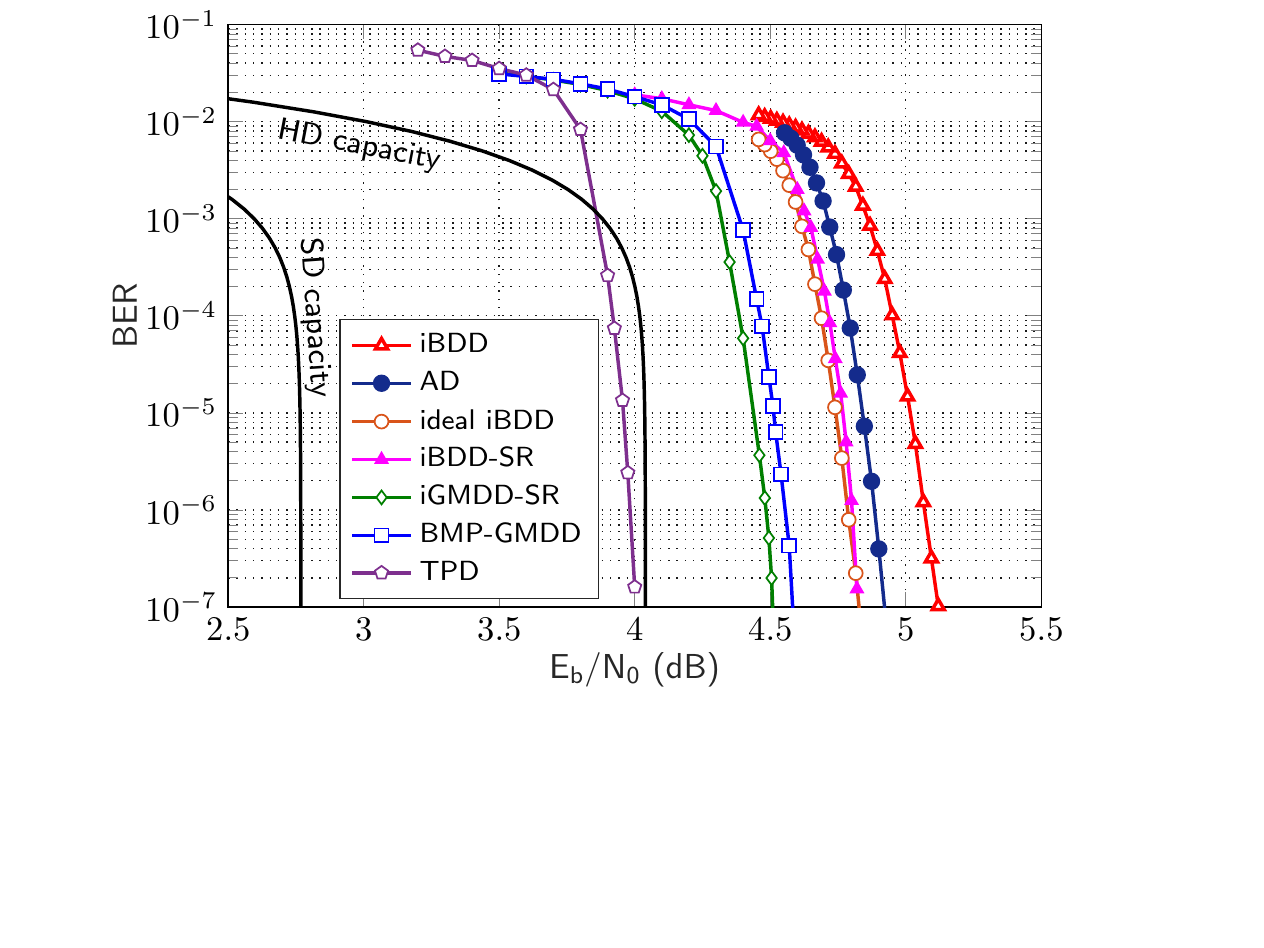}  
	\caption{BER performance of different decoding algorithms for a PC with  ($256,239,6$) eBCH component codes and transmission over the AWGN channel. The PC rate is $0.8716$ and the maximum number of decoding iterations is $10$.} 
	\label{v9t4s7_product} 
\end{figure}

In Fig.~\ref{v9t4s7_product}, we give the bit error rate (BER) performance of BMP-GMDD  for a PC with
double-error-correcting extended BCH (eBCH) codes with parameters
$(256,239,6)$ as component codes and transmission over the AWGN channel. The resulting PC has rate 
$R = 239^2/256^2 \approx 0.8716$.  For comparison purposes, we we also plot the performance of conventional iBDD, AD \cite{Hag18}, iBDD-SR \cite{She18}, iGMDD-SR \cite{She18b}, and TPD based on the Chase-Pyndiah decoding \cite{Pyn98}. For all algorithms, a maximum of $\maxIter =
10$ decoding iterations is performed. As a reference, the Shannon limit for SDD and HDD is also plotted in the figure.

Both BMP-GMDD and iGMDD-SR require a proper choice of the scaling
factors $w_i^{(\ell)}$. For simplicity, we consider the same scaling factor for all row and column codes, i.e., $w_i^{\mathsf r, (\ell)} = w_j^{\mathsf c, (\ell)}= w^{(\ell)}$ for all $i,j=1,\ldots,n$, and jointly optimize the vector $\w = (w^{(1)}, \dots, w^{(\maxIter)})$ by using Monte--Carlo
estimates of the BER for a fixed $\Eb/\No$ as the
optimization criterion. Intuitively, one would expect that the
decisions of the component decoders become more reliable with
increasing number of iterations, whereas the channel observations become less informative.
Therefore, in order to reduce the optimization search space, we only consider vectors $\w$ with monotonically increasing entries.  iBDD-SR also requires scaling factors (see \cite{She18,She19}). In this case, the scaling factors can be derived using density evolution \cite{She19,lechner2012analysis}.

The two reference curves are conventional iBDD (red curve with empty triangle markers) and TPD (purple curve with pentagon markers), with the latter performing  $1.1$ dB better at a BER of $10^{-7}$. AD (dark blue curve with filled circle markers) and iBDD-SR (pink curve with filled triangle markers) outperform  conventional iBDD by $0.18$ dB and
$0.27$ dB, respectively, at the same BER. As a
reference, we also show the performance of \emph{ideal} iBDD (brown curve with empty circle markers), where a genie
prevents miscorrections. Interestingly, at a BER of $10^{-7}$ iBDD-SR yields the same performance as ideal iBDD.\footnote{We remark that the performance of iBDD-SR in Fig.~\ref{v9t4s7_product} is improved compared to \cite{She18b}, since in this paper we use the optimized scaling factors based on the density evolution derived in \cite{She19}, rather than based on Monte-Carlo simulations as in \cite{She18b}.}  
iGMDD-SR (green curve with diamond markers) outperforms iBDD, iBDD-SR, and AD and closes $\approx 54\%$ of the gap between iBDD and TPD, at the expense of an increased complexity and data flow. 

The performance of the proposed BMP-GMDD is given by the blue curve with square markers. The proposed decoding algorithm yields performance very close to that of iGMDD-SR (a performance degradation compared to iGMDD-SR of only $0.074$ dB is observed at a BER of $10^{-7}$), while achieving a significantly lower data flow.
BMP-GMDD closes around $50$\% of the performance gap between iBDD and TPD, while requiring only a $21.48\%$ higher data throughput than iBDD.

The coding gain improvements of all considered decoding algorithms
over iBDD are summarized in Table~\ref{tab:Table1} (fifth column). In the table we also indicate whether the algorithms exploit the channel reliabilities or not, the nature of the messages exchanged in the iterative decoding (hard or soft), as well as the gap to
capacity for all schemes (sixth column). Note that the performance of iBDD and AD
should be compared to the hard decision (HD) capacity, while the performance of
iBDD-SR, iGMDD-SR, BMP-GMDD, and TPD should be compared to the soft decision (SD) capacity since the channel LLRs are exploited in the decoding. Overall, one can see a
clear trade-off between BER performance and decoding complexity for the different
algorithms.


We remark that if the channel LLRs are highly reliable but with wrong sign, one can expect that the decoding rule in \eqref{eq:GMDchrel_VN_scale} will be unable to recover from these errors. In this situation, although $\bar{\mu}_{i,j}^{\mathsf r}$ may correspond to a correct decision, it is overridden by the channel channel, i.e., the hard decision on code bit $c_{i,j}$ made by the $i$-th row decoder, $\psi_{i,j}^{\mathsf{r},(\ell)}$, becomes  $\psi_{i,j}^{\mathsf{r},(\ell)}=\BB(w_i^{\mathsf r, (\ell)} \cdot \bar{\mu}_{i,j}^{\mathsf r,(\ell)} + L_{i,j})=\BB(L_{i,j})$ (cf. \eqref{eq:GMDchrel_VN_scale} and \eqref{eq:BDDchrel_VN_scale}), leading to an erroneously decoded bit. Therefore, one needs to be careful when applying BMP-GMDD to avoid the appearance of an error floor. In particular, to avoid such errors and avoid a high error floor, we run BMP-GMDD for some iterations and then we append a few conventional iBDD iterations, where the channel reliabilities are disregarded when making the decision on a given code bit. The appended iBDD iterations increase the chance to correct transmission errors with high channel reliability. By doing so, an error floor is avoided. The same discussion applies to iBDD-SR \cite{She18,She19} and iGMDD-SR \cite{She18b}. For the simulation of BMP-GMDD, iBDD-SR, and iGMDD-SR in Fig.~\ref{v9t4s7_product}  we considered $8$ decoding iterations of the algorithms appended with $2$ iBDD iterations. 

\section{Conclusion}
\label{conclusion}

We proposed a new message passing decoding algorithm for product codes based on generalized minimum distance decoding, i.e., error and erasure decoding, of the component codes, where the last stage of GMDD is based on the Hamming distance metric. The proposed algorithm, dubbed BMP-GMDD, exploits soft information but requires to exchange only hard decisions and a short ordered list of the least reliable bits between component decoders, hence introducing a limited increase in data flow compared to conventional iterative bounded distance decoding. For the considered scenario based on $(256,239,6)$ double-error-correcting eBCH component codes, the proposed algorithm closes about $50\%$ of the performance gap between iBDD and turbo product decoding and yields performance very close to that of the algorithm iGMDD-SR introduced in \cite{She18b}, with a much lower data flow, only $21.48\%$ higher than that of iBDD. The increase in data flow is even lower for longer component codes. 
While in this paper we considered PCs, the proposed algorithm can be extended to other classes of product-like codes such as staircase codes. Overall, the proposed BMP-GMDD algorithm provides a very good performance-complexity tradeoff and is appealing for very high-throughput applications such as fiber-optic communications.

\section*{Acknowledgment}

The authors would like to thank Dr. Christian H\"ager for providing the simulation results of anchor decoding in Fig.~\ref{v9t4s7_product}.

\end{document}